\documentclass{lmcs} 

\usepackage{hyperref}

\usepackage{fancyvrb}
\usepackage{graphicx}%
\usepackage{multirow}%
\usepackage{amsmath,amssymb,amsfonts}%
\usepackage{amsthm}%
\usepackage{mathrsfs}%
\usepackage[title]{appendix}%
\usepackage{xcolor}%
\usepackage{textcomp}%
\usepackage{manyfoot}%
\usepackage{booktabs}%
\usepackage{algorithm}%
\usepackage{algorithmicx}%
\usepackage{algpseudocode}%
\usepackage{listings}%

\makeatletter
\newenvironment{breakablealgorithm}
  {% \begin{breakablealgorithm}
    \begin{flushleft}
    
      \refstepcounter{algorithm}% New algorithm
      \hrule height.8pt depth0pt \kern2pt% \@fs@pre for \@fs@ruled
      \parskip 0pt
      \renewcommand{\caption}[2][\relax]{% Make a new \caption
        {\raggedright\textbf{\fname@algorithm~\thealgorithm} ##2\par}%
        \ifx\relax##1\relax % #1 is \relax
          \addcontentsline{loa}{algorithm}{\protect\numberline{\thealgorithm}##2}%
        \else % #1 is not \relax
          \addcontentsline{loa}{algorithm}{\protect\numberline{\thealgorithm}##1}%
        \fi
        \kern2pt\hrule\kern2pt
     }
  }
  {% \end{breakablealgorithm}
     \kern2pt\hrule\relax% \@fs@post for \@fs@ruled
    \end{flushleft}
  }
\makeatother

\theoremstyle{plain} %\crefname{satz}{Satz}{S\"atze}

\begin{document}

\title[Type Inference for Isabelle2Cpp]{Type Inference for Isabelle2Cpp}
\author[J.~Dongchen]{Jiang Dongchen}[a]
\author[F.~Chenxi]{Fu Chenxi}[b]

\address{School of Information Science and Technology, Beijing Forestry University, Beijing $100083$, China.}	%optional
\email{jiangdongchen@bjfu.edu.cn}

\address{School of Information Science and Technology, Beijing Forestry University, Beijing $100083$, China.}	%optional
\email{steve7800@bjfu.edu.cn}

\begin{abstract}
  \noindent Isabelle2Cpp is a code generation framework that supports automatic generation of C++ code from Isabelle/HOL specifications. However, if some type information of Isabelle/HOL specification is missing, Isabelle2Cpp may not complete the code generation automatically. In order to solve this problem, this paper provides a type system for Isabelle2Cpp, which is used to perform type inference and type unification for expressions of the intermediate representation in Isabelle2Cpp. The system introduces new type inference rules and unification algorithms to enhance the Isabelle2Cpp framework. By incorporating the type system, the Isabelle2Cpp framework can provide more comprehensive type information for expression generation, which leads to more complete and accurate C++ code.
\end{abstract}

\maketitle

\section{Introduction}\label{S:one}

  Isabelle/HOL is a general interactive theorem proof assistant that provides a formal environment for specification writing and property proving\cite{nipkow2002isabelle}. It has been widely used for mechanical proofs of various mathematical theorems, and for correctness verification of algorithms, software and some complex systems. Users can define data types and function specifications, and complete proofs for relevant properties.

After proving the properties of related specifications, a direct idea to follow is to obtain executable codes from the specifications automatically. Code generation is the technique used for generating executable codes based on given specifications, and it establishes the equivalent semantic transformation from specifications to programs of the target language. The generated codes inherit the verified properties of the original specifications, which increases the application scenarios of the specifications.

Isabelle2Cpp is an automatic conversion framework that allows to generate C++ implementation from Isabelle/HOL specifications\cite{jiang2022generation}. This framework combines the verification convenience of Isabelle/HOL and the execution efficiency of C++, and separates the C++ code generation away from the property verification of Isabelle/HOL specifications. Since the correctness of the Isabelle/HOL specification can be guaranteed by interactive proof, the correctness of the generated C++ codes can also be maintained.

As Isabelle/HOL and C++ use different type inference systems, once the type information of an Isabelle/HOL specification is incomplete, the types of some variables in the generated C++ codes can not be inferred by using the auto-deltype-based type reasoning mechanism in C++. It still requires manual completion of the missing types in the partially generated C++ codes. Therefore, the purpose of this paper is to provide the type system for Isabelle2Cpp, which is used for type inference and type instantiation for C++ code generation from Isabelle/HOL specifications.

This paper is organized as follows: Section 2 reviews the relevant work. Section 3 describes the Isabelle2Cpp framework and analyzes the problems in type inference. Section 4 provides the type inference system of Isabelle2Cpp, and Section 5 provides its algorithm implementations. Experiment and case study are presented in Section 6, while Section 7 summarizes the whole paper.

\section{Related Works}\label{S:two}
Type systems plays an important role in the design of programming languages and type-based program analysis. In programming, a type system is a logical system composed of a set of rules that assigns a type to each term (a variable, expression or function)\cite{cardelli1996type}. A type system is often specified as part of a programming language, and the process of type inference allows programmers to omit type annotations while still enabling type checks, and ensures that the program is well-typed\cite{duggan1996explaining}.

One of the early well-known type systems is System F, which is an extension of the lambda calculus with polymorphic types\cite{girard1986system}. It was independently discovered by Jean-Yves Girard and John C. Reynolds in 1971 and officially named System F by Reynolds in 1972. One key feature of System F is that it introduces type abstraction and type application, making the types themselves first-class citizens, i.e. types can be passed as parameters to functions and as return values of functions. This feature enables System F to support higher-order polymorphism.

Another well-known type systems is the Hindley-Milner (HM) type system\cite{milner1978theory}. It was initially described by J. Roger Hindley and later rediscovered by Robin Milner\cite{damas1984type}, with detailed formal analysis and proof provided by Luis Damas in his doctoral thesis\cite{damas1982principal}. The main difference between the HM type system and System F is the way polymorphism is expressed: The HM type system is simple and can achieve local type polymorphism by performing type inference on function parameters and return values. In contrast, System F supports global type polymorphism, which makes it has higher expressive power and can express more complex polymorphism patterns.

HM type system supports parametric polymorphism and was firstly implemented as part of the ML programming language, and has been extended in various languages.
One notable feature of the HM type system is its ability to infer the most general type of a given program without type annotations or other hints from the programmer. Milner's algorithm W is an effective type inference method in practice, which has been successfully applied to large codebases. Another type inference algorithm for ML is M algorithm whose reliability and completeness were proven by Lee Oukseh and Yi Kwangkeun\cite{lee1998proofs}. Being different from algorithm W, algorithm M is a top-down algorithm, which can detect type errors earlier. Duggan Dominic and Bent Frederick modified the unification algorithm for the HM type inference, allowing specific inferences for type interpretation.

Soosaipillai Helen combined the idea of interpretive reasoning and proposed a method for type checking of SML programs\cite{soosaipillai1990explanation}. Specifically, the type checker generates interpretations from existing types, which are used to infer interpretations of unknown types, and then infer the specific types of type variables. This method improves the efficiency of type checking.
Heeren BJ, Hage Jurriaan and others described an approach that generates a set of type constraints for expressions\cite{heeren2002generalizing}. These constraints are typically generated locally so that the framework can be applied to more complex type inference problems.
Apart from the HM type system, Kfoury Assaf J and Wells Joe B defined an intersection type system with primary types and strongly normalizable $\lambda$-term types\cite{kfoury2004principality}. The system can handle complex cross-type inference and ensure the uniqueness and consistency of type inference.
R{\'e}my Didier considered pure function semantics and call-by-value operational semantics to specify higher-level polymorphism through type annotations on source terms, allowing implicit predicate types and explicit predictive type instantiation\cite{remy2005simple}. 
Vytiniotis Dimitrios, Jones Simon Peyton and others proposed a constraint-based local type inference method based on the OutsideIn(X) system, whose parameterization is similar to HM(X) on a specific foundation constraint domain X\cite{vytiniotis2011outsidein}. This constraint solver has been implemented as part of GHC 7.
Traytel Dmitriy, Berghofer Stefan andothers added coercive subtyping to the type system of a simple typed lambda calculus with Hindley-Milner polymorphism and proposed a type inference algorithm that determines where to insert coercive subtyping to ensure correct typing based on the HM type system\cite{traytel2011extending}. The algorithm is reliable and complete, and has been implemented in Isabelle.

In the domain of non-functional languages, Palsberg Jens and Schwartzbach Michael I proposed a method for type inference in non-object-oriented programs that include inheritance, assignments, and late binding\cite{palsberg1991object}. The method guarantees that all messages are understood, annotates the program with type information, allows for polymorphic methods, and serves as the basis for optimizing compilers.
Motivated by statistical analysis of type inference in ML codebases, Pierce Benjamin C and Turner David N explored two partially type inference methods for languages that combine subtyping and imperative polymorphism\cite{pierce2000local}. One method uses a local constraint solver to infer type parameters in polymorphic applications, while the other infers annotations for binding variables in function abstractions by propagating type constraints downward from enclosing application nodes. 
For type inference in stripped binary files, Chen Ligeng, He Zhongling and others proposed the CATI algorithm\cite{chen2020cati}, which captures the context of function calls during execution, and infers the types of function arguments and return values.
Traditional rule-based type inference methods usually rely on the syntax rules and type systems of programming languages, which limits the applicability to complex code structures and dynamically typed languages to a certain extent. In response to the limitations of traditional methods, Hellendoorn Vincent J, Bird Christian and others proposed a method based on deep learning to implement the code type inference system DeepTyper \cite{hellendoorn2018deep}. The core idea of DeepTyper is to infer the type of code by learning the structure and semantic information of code fragments, rather than relying on static rules and type systems.

The type system of Isabelle/HOL is the standard HM type system, which is different from the auto-deltype-based type reasoning mechanism in C++. This difference makes type inferences of some variables in the generated C++ codes (by Isabelle2Cpp) and the corresponding terms in Isabelle/HOL specifications may be inconsistent. As the type system of Isabelle/HOL can not be used for the type inference of generated C++ codes directly, it is necessary to provide a corresponding type system for Isabelle2Cpp.

Compared to existing works, the type system proposed in this paper is based on the HM type system and is specifically designed for the Isabelle2Cpp code generation framework. It introduces new type inference rules and unification algorithms to enhance the Isabelle2Cpp framework by adding a type system and completing type inference. We made our efforts to minimize modifications to the standard HM type system but avoided introducing additional concepts such as constraint solving or coercion to reduce system complexity. By incorporating our type inference algorithm, the Isabelle2Cpp framework can provide more comprehensive type information for expression generation, enabling the generation of more complete and accurate C++ code.
\section{Framework of Isabelle2Cpp}\label{S:three}
The architecture of Isabelle2Cpp is shown in Fig. 1. For a given specification in Isabelle/HOL, a standard intermediate representation abstract syntax tree (AST) is constructed by syntax parsing firstly, and it contains all semantic information of the specification; secondly, all information stored in AST, e.g. specification names, type variables, datatypes and their corresponding construction rules, and equations to define functions, are used for core part generation of the C++ codes, and all conceptional generations are done in the conversion part; lastly, the synthesis part will produce concrete C++ files by adding the corresponding header file so that program can be compiled and executed directly.

\begin{figure}
\centerline
  {\includegraphics[scale=1.0]{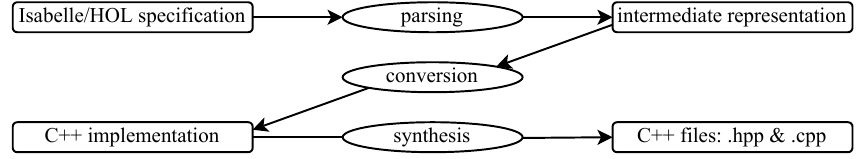}}\vskip3mm

\caption{\label{fig:1} Architecture of Isabelle2Cpp.}

\end{figure}

The Isabelle2Cpp framework supports two ways for conversion: the definition-based conversion and the rule-based conversion.

The definition-based method converts Isabelle/HOL datatypes into C++ classes or class templates where each construction rule of a datatype is expressed by a C++ nested class, and converts Isabelle/HOL functions into sequences of semantically equivalent statements in the generated C++ functions or function templates where each equation of a Isabelle/HOL function is converted into an if-statement block (called a conditional statements module or CSM for short) in C++. In the function conversion, the pattern of an equation is converted into conditions and optional declarations, and the expression is converted into a final return-statement and some optional auxiliary statements.
The correctness of the definition-based conversion is guaranteed by the facts that:
1) the conversion mapping from Isabelle/HOL specification to C++ code is isomorphic;
2) the basic operations of the Isabelle/HOL specification are semantically equivalent to the basic statements in CSM of the C++ codes.

The rule-based conversion is also provided in Isabelle2Cpp to improve the efficiency of the generated C++ codes. Rules for datatype and operation conversions are manually provided, thus existing predefined types and relevant operations in C++ can be directly utilized in code generation. Usually, the corresponding entities (datatypes/types or functions) in a conversion rule should express the same mathematical concepts in the two languages. As the structures of the entities of a conversion rule may be different in two language, the correctness of this conversion method can not be totally guaranteed. However, if the conversion rules, target types and functions are carefully defined, the correctness of the generated C++ codes can also be assured. Generally, the rule-based conversion is more flexible and the generated C++ codes are more efficient.

\subsection{Problems}
Isabelle2Cpp will automatically generate executable C++ codes if the Isabelle/HOL specification contains all necessary information, e.g. type names, type variables and construction rules of type definitions, function names, function types and equations of function definitions [see Fig. 2]. However, as type inference is supported by Isabelle/HOL, some type information can be omitted in Isabelle/HOL specifications without errors, but this will lead to type information missing in ASTs and make the generated C++ codes can not be compiled directly.

\begin{figure}
\centerline
  {\includegraphics[scale=0.8]{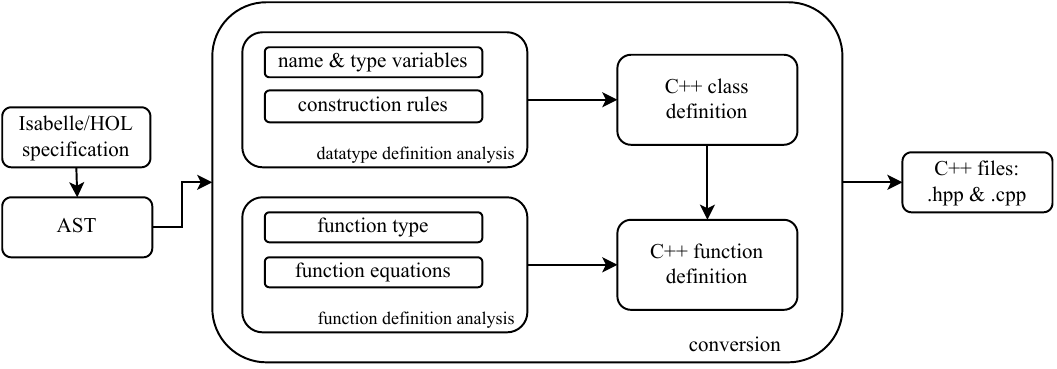}}\vskip3mm

\caption{\label{fig:2} Conversion process of Isabelle2Cpp.}

\end{figure}

In order to solve the problem of type missing in generated C++ codes, Jiang and Xu used the $\mathbf{decltype}$ specifier to inspect the type of an entity [3]. To determine the type of newly declared variables, the framework first checks whether the type information is provided in AST: if no type information is not found, the unlabeled variables are specified with $\mathbf{auto}$, then the specific data type of an $\mathbf{auto}$ variable can be inferred according to the existing type information of its relevant expression by using $\mathbf{decltype}$ in C++.

\begin{Verbatim}[frame=single, fontsize=\small]
fun test :: "'a list => nat" where
  "test Nil = 0" |
  "test (Cons x xs) = length (If ((length xs) = 0) Nil xs) + 1"
\end{Verbatim}

\begin{Verbatim}[frame=single, fontsize=\small]
std::uint64_t test(const std::list<T1> &arg1) {
    ...
    // test (Cons x xs) = length (If ((length xs) = 0) Nil xs) + 1
    if (!arg1.empty()) {
        auto xs = decltype(arg1){std::next(arg1.begin()), arg1.end()};
        UNKNOWN_TYPE temp0;
        if (xs.size() == 0) {
            temp0 = {};
        } else {
            temp0 = xs;
        }
        return temp0.size() + 1;
    } else { // auto-generated for -Wreturn-type
        std::abort();
    }
}
\end{Verbatim}

Take the Isabelle/HOL function $test$ as an example: Without a type inference system, Isabelle2Cpp can not infer  $xs$ with a type of ${'a}\ list$, thus the variable $xs$ is specified as auto in C++. As the type of $xs$ is the same as the type of $(Cons\ x\ xs)$, which is the type of the first parameter $arg1$ in the generated C++ function, the type of $xs$ can be obtained through $\mathbf{decltype}$, i.e. $decltype (arg1) \{std::next (arg1.begin()),\ arg1.end()\}$.

$\mathbf{decltype}$ is useful to inspect types of specific expressions at the C++ code level, e.g.  lambda-related types or types that depend on template parameters. However, obtaining type information at the C++ code level still can not solve the following problems for code generation:

Firstly, $\mathbf{decltype}$ cannot solve the problem of type unification in Isabelle2Cpp. For example, in the specification of $test$, $If$ can be seen as a function with a type $bool \Rightarrow {'a} \Rightarrow {'a} \Rightarrow {'a}$. Isabelle2Cpp converts the $If$ expression into a if-statement block with an additional temporary variable $temp0$. According to the datatype of $If$ expression in Isabelle/HOL, the types of its second and third parameters, as well as the return value type of whole expression, are the same. However, due to the lack of type unification tool, Isabelle2Cpp is unable to infer that the type of the second parameter $Nil$ is the same with the type of the third parameter $xs$ from the type of $If$ expression. Although the type of $xs$ can be specified by $\mathbf{decltype}$ during compilation, the type of $temp0$ cannot be determined during code generation and can only be denoted by UNKNOWN\_TYPE which need to be specified manually afterwards.

Secondly, Isabelle2Cpp convert Isabelle/HOL lambda expressions into C++ lambda expressions directly. In Isabelle/HOL specification, a lambda expression can not only be used as a function call, but also be passed to a higher-order function as a parameter. For the latter, Isabelle2Cpp uses std:: function class template to represent the type of function parameters, which should be instantiated by the type of corresponding lambda expression. Due to the fact that C++ lambda expressions are implemented as anonymous structures, inconsistency will occur if a C++ lambda expression is set as a parameter of a higher-order function.

\begin{Verbatim}[frame=single, fontsize=\small, commandchars=\\\{\}, codes={\catcode`$=3\catcode`_=8}]
[&] (T1 x$_1$, ..., Tn x$\rm_n$) \{ return t$\rm_c$; \}
\end{Verbatim}

Take the $map$ function as an example: Isabelle2Cpp generates the following function template for $map$ in C++ where the type of its first parameter is std::function\textless R(T)\textgreater. If we declare an auto variable $f$ to receive a lambda expression in the main function, its type will be derived as class lambda [] (auto\ x) $\rightarrow$ auto, which is inconsistent with the parameter type std::function\textless R(T)\textgreater. But if  the specific type of a lambda expression can be derived before the code generation, then the type of the std::function class template in a higher-order function can be automatically instantiated. For example, if we know the parameter type and the return value type of a lambda expression are $nat$, then a std::function\textless int\textless int\textgreater \textgreater\ type variable will be used to receive the lambda expression.

\begin{Verbatim}[frame=single, fontsize=\small]
template<typename T, typename R>
std::vector<R> map(std::function<R(T)> f, std::vector<T> xs) {
    auto impl = [&]() -> R {
        //map f [] = []
        if (xs.empty()) {
            return R();
        }
        //map f (x : xs) = (f x) : (map f xs)
        auto x = xs.front();
        xs.erase(xs.begin());
        return R({f(x)}) + map(f, xs);
    };
    static std::map<std::tuple<std::function<R(T)>, std::vector<T>>, R> cache;
    auto args = std::make_tuple(f, xs);
    auto it = cache.find(args);
    return it != cache.end() ? it->second : (cache.emplace(std::move(args),
        impl()).first->second);
}

int main() {
    auto f = [](auto x) -> auto { return x * x; };
    auto xs = std::vector<std::uint64_t>({1, 2, 3, 4, 5, 6, 7, 8, 9, 10});
    auto ys = map(f, xs);
    for (auto y : ys) {
        std::cout << y << std::endl;
    }
}
\end{Verbatim}

The above examples have shown that type inference at the C++ code level alone cannot eliminate the problems caused by type information missing in Isabelle/HOL specifications. It is necessary to provide a type system for the framework of Isabelle2Cpp to make type inference and totally automatic generation of C++ codes from executable Isabelle/HOL specifications.

\section{Type System of Isabelle2Cpp}\label{S:four}
    
\subsection{Function specifications in Isabelle/HOL}

In Isabelle/HOL, a function is defined by providing its name, type and the relationship between its inputs and output, and the specification satisfies the following syntax \cite{krauss2008defining}:
\begin{align*}
&{\bf fun}\ f\ {::}\ ''t_{\rm 1}\Rightarrow...\Rightarrow t_{\rm N}\Rightarrow t_{\rm R}''\ {\bf where}\\
&\ \ ''pattern_{\rm 1} = expression_{\rm 1}''\\
&\ \ \vdots\\
&\ \ ''pattern_{\rm n} = expression_{\rm n}''
\end{align*}

Here, $f$ is the function name, $t_1$, ..., $t_{\rm N}$ are the types of its parameters, $t_{\rm R}$ is the type of return expression, and $pattern_{\rm i} = expression_{\rm i}\ ({\rm i}=1,\dots,n)$ are the equations to define $f$ where pattern matching is used. On the left side of an equation, a pattern is represented by a datatype constructor with relevant pattern parameters; on the right side, an expression to define the function with the corresponding pattern is provided.

In the conversion process of Isabelle2Cpp, all necessary information for code generation should be parsed and stored in AST. As some function specification may not directly provide all detailed type information about variables and expressions, it is necessary to store existing type information for type inference. Thus, the structure of AST is modified where type nodes are added for all pattern parameters and expressions (including the corresponding sub-expressions) [see Fig. 3].
\begin{figure}[H]
\centerline
  {\includegraphics[scale=0.3]{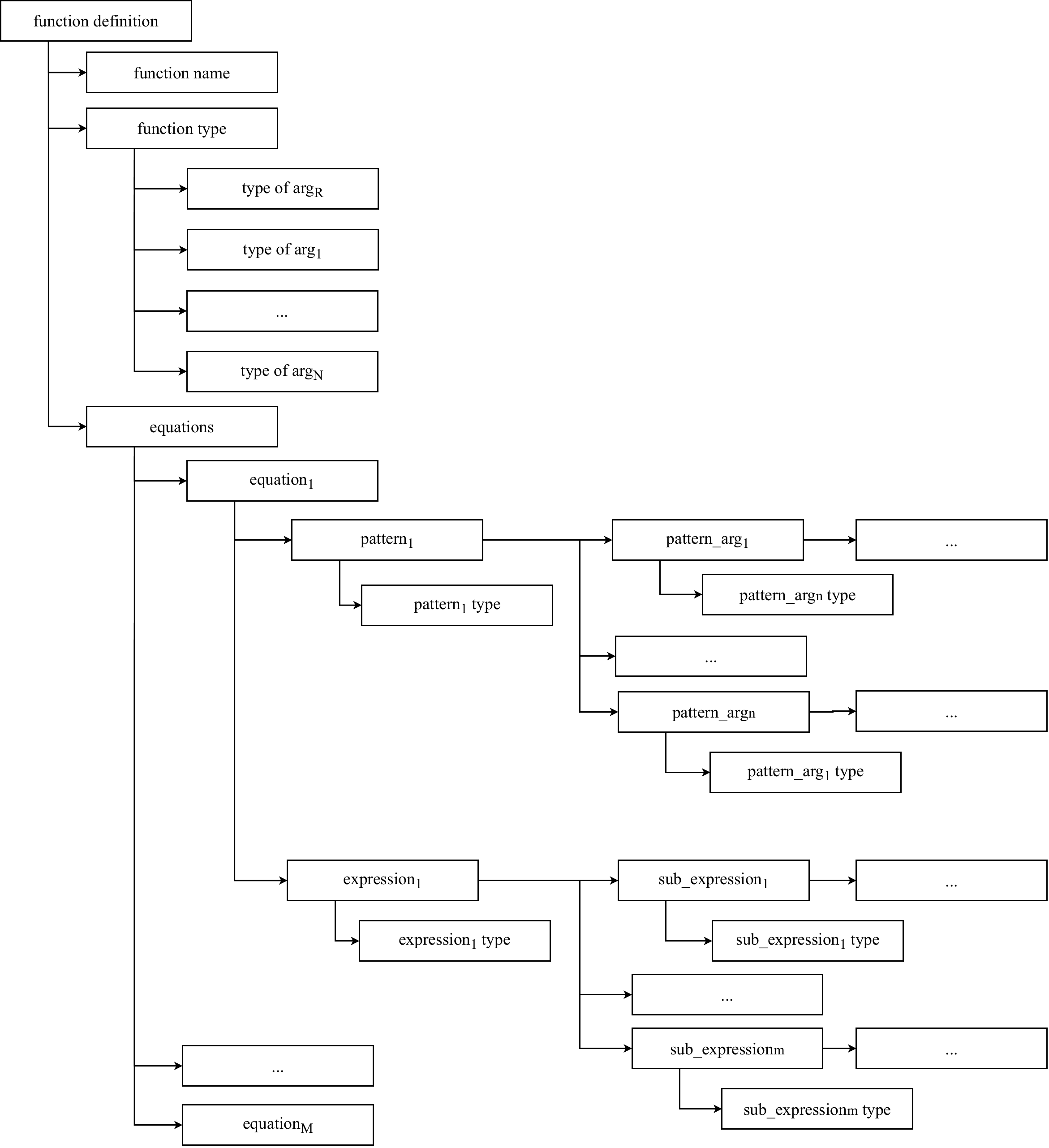}}\vskip3mm

\caption{\label{fig:3} An abstract syntax tree with type information variable type.}

\end{figure}

\subsection{Expressions \& Type Constructions in Isabelle2Cpp}
To define the type system of Isabelle2Cpp, expressions in AST should be formally defined: 
\vspace{-10pt}
\begin{align*}
e\ :=&\ x\ |\ (C\ e_1\ e_2\ \cdots \ e_n)\ |\ (if\ e\ then\ e'\ else\ e'')\ |\ (\lambda\ x_1\ \cdots \ x_n.\ e)\ |\ \\
&(case\ e\ of\ pattern_1\ =>\ e_1\ |\ pattern_m\ =>\ e_m)\ |\ \\
&(let\ x\ =\ e\ in\ e')\ |\ [e_1\ \cdots\ e_n]\ |\ \{e_1\ \cdots\ e_n\}|\ \cdots\ 
\end{align*}

An expression in AST can be a constant expression (e.g. $1,2,True,False$), a variable expression, an application expression, an if-expression, a lambda-expression, a case expression, a let-in expression, a list expression or a set expression. It should be noted that function application expressions and value construction expressions are both regarded as application expressions in this definition as an n-ary function and an n-ary value construction have similar characteristics: once n values with proper types are provided, a new value will be obtained by the function application or the value construction. In the definition, a list expression or a set expression can be regarded as a special case of a function application expression. They are specified because the rule-based conversion of Isabelle2Cpp supports the direct conversion for the corresponding type constructors.

To provide type information for expressions, type expressions are defined as follows: 
$$
\tau:=\alpha\ |\ T\ |\ (\tau)\ |\ \tau_1 \Rightarrow \tau_2\ |\ (\tau_1,\ \tau_2)\ |\ \tau_1 \dots \tau_n\ D\ |\ 'a\ list\ |\ 'a\ set\ \dots
$$\vspace{0pt}
i.e. a type expression can be a type variable $(\alpha, \beta, etc.)$, a primitive type $(bool, nat, etc.)$, a function type, a tuple type, a new type constructed by an abstract type constructor with existing types, a list type or a set type. 

Once the concept of types is provided, the type variable set $Var(\tau)$ of a given type $\tau$ can be inductively defined, i.e.
\begin{defi}
For a given type $\tau$,
\begin{itemize}
  \item if $\tau$ is a type variable $\alpha$, then $Var(\tau)=\{\alpha\}$;
  \item if $\tau$ is a primitive type $T$, then $Var(\tau)=\emptyset$;
  \item if $\tau=(\tau_1)$, then $Var(\tau)=Var(\tau_1)$;
  \item if $\tau=\tau_1\Rightarrow\tau_2$, then $Var(\tau)=Var(\tau_1)\cup Var(\tau_2)$;
  \item if $\tau=(\tau_1,\tau_2)$, then $Var(\tau)=Var(\tau_1)\cup Var(\tau_2)$;
  \item if $\tau=\tau_1\dots\tau_n~D$, then $Var(\tau)=Var(\tau_1)\cup\dots\cup Var(\tau_n)$;
  \item if $\tau=\alpha~list$, then $Var(\tau)=\{\alpha\}$;
  \item if $\tau=\alpha~set$, then $Var(\tau)=\{\alpha\}$.
\end{itemize}
\end{defi}

\subsection{Rules of Type Inference in Isabelle2Cpp}

In Isabelle2Cpp, if $e$ is an expression and $\tau$ is a type expression, a typing (judgement) $e :\tau$ is an assertion that the type of $e$ is $\tau$.
A type context $\Gamma$ is a mapping from variables to types, which can be represented as a form of $\{v_1: \tau_1,\dots,\ v_n: \tau_n\}$. A typing formula is a formal statement of the form $\Gamma \vdash e : \tau$ where $\Gamma$ is a type context, $e$ an expression and $\tau$ a type. A derivable typing is a typing that can be inferred from a consistent type context by the type system.
The inference rules of the type system in Isabelle2Cpp are divided into three parts according to their functionality: the top-down rules, the bottom-up rules, and the unification rules.

The top-down rules are mainly used to obtain the type information of parameters in a function by following a top-down type completion approach. In Isabelle/HOL, functions are defined by pattern matching. Therefore, the type system can obtain the general type information of function parameters from the specification and infer the detailed information of the actual pattern parameters based on the type of corresponding abstract type constructors. Thus, the top-down rules for abstract type constructor are given by $\mathbf{App-TD}$, $\mathbf{List-TD}$ and $\mathbf{Set-TD}$.
Besides, as the type of the return expression is given in the specification, the top-down rules can also be used in type instantiation for all sub-expressions. Thus, the rules $\mathbf{Let-TD}$ and $\mathbf{Case-TD}$ for other expression constructions are also provided. 

%TD
\begin{equation*}
\boxed{
\begin{gathered}
\frac{\Gamma \vdash [e_1, \cdots e_n]:\tau\ list}{\Gamma \vdash e_1:\tau, \cdots e_n:\tau}\quad[\mathbf{List-TD}] \\
\frac{\Gamma \vdash \{e_1, \cdots e_n\}:\tau\ set}{\Gamma \vdash e_1:\tau, \cdots e_n:\tau}\quad[\mathbf{Set-TD}] \\
\frac{\Gamma \vdash C \ e_1\ \cdots  e_n : \tau_r\quad \sum(C)= \tau_1 \Rightarrow \cdots \Rightarrow \tau_n \Rightarrow \tau_r}{\Gamma \vdash \ e_1:\tau_1,\cdots, e_n:\tau_n }\quad[\mathbf{App-TD}] \\
\frac{\Gamma \vdash \mathbf{let} \ x = e_0 \ \mathbf{in}\ e_1 : \tau}{\Gamma \vdash e_1 : \tau} \quad[\mathbf{Let-TD}] \\
\frac{\Gamma \vdash (\mathbf{Case}\  e\ \mathbf{of}\  pattern_1\Rightarrow e_1\ \cdots\  pattern_m\Rightarrow e_m):\tau}{\Gamma \vdash e_1 : \tau \cdots, e_m : \tau \quad}\quad [\mathbf{Case-TD}] \\
\end{gathered}
}
\end{equation*}

In this paper, a type solver $\Sigma$ is used to record the type information of functions or datatype constructors (e.g. $\Sigma (Cons) = {'a} \Rightarrow {'a}\ list \Rightarrow {'a}\ list$ indicates that the abstract type constructor $Cons$ has the type of\ ${'a} \Rightarrow {'a}\ list \Rightarrow {'a}\ list$, and $\Sigma (length)={'a}\ list\Rightarrow nat$ indicates that the function length has the type of ${'a}\ list \Rightarrow nat$).

The bottom-up rules are used to derive the type for a whole expression from the types of its immediate subexpressions, and they are defined as follows:

\begin{equation*}
\boxed{
\begin{gathered}
\frac{e:\sigma \in \Gamma}
{\Gamma \vdash e:\sigma }
\quad[\mathbf{Exp-BU}] \\
\frac{\Gamma \vdash  e_1 : \tau_1,\cdots\ e_n:\tau_n \quad \sum(C)=\tau_1 \cdots \tau_n \Rightarrow \tau_r}
{\Gamma \vdash \ C\ e_1\ \cdots\ e_n : \tau_r}
\quad [\mathbf{App-BU}] \\
\frac{\Gamma \vdash e_0 : \sigma \quad \Gamma,x:\sigma \vdash e_1 : \tau}
{\Gamma \vdash \mathbf{let} \ x = e_0 \ \mathbf{in}\ e_1 : \tau}
\quad[\mathbf{Let-BU}] \\
\frac{\Gamma \vdash e : \tau \quad \Gamma \vdash e_1 : \tau' \cdots \Gamma \vdash e_m : \tau' \quad}
{\parbox{9cm}{${\Gamma \vdash (\mathbf{Case}\  e \ \mathbf{of}\  pattern_1 \Rightarrow e_1 \cdots\  pattern_m \Rightarrow e_m):\tau'},\\ \ {\hphantom{aaaaaaaaaa}\Gamma \vdash pattern_1:\tau \cdots pattern_m:\tau}$}}
\quad [\mathbf{Case-BU}] \\
\frac{\Gamma, x_1:\tau_1, \dots, x_n:\tau_n \vdash e : \tau'}
{\Gamma \vdash \lambda x_1, \dots x_n.e:\tau_1 \Rightarrow \dots \Rightarrow \tau_n \Rightarrow \tau'}
\quad[\mathbf{Abs-BU}] \\
\frac{\Gamma \vdash e_1:\tau,  \cdots, e_n:\tau}
{\Gamma \vdash [e_1, \cdots, e_n]:\tau\ list}
\quad[\mathbf{List-BU}] \\
\frac{\Gamma \vdash e_1:\tau,\cdots, e_n:\tau}
{\Gamma \vdash \{e_1, \cdots, e_n\}:\tau\ set}
\quad[\mathbf{Set-BU}]
\end{gathered}
}
\end{equation*}

Among the above rules, $\mathbf{Exp-BU}$ is the atomic rule which shows that the existing typing can be inferred directly from its context. Furthermore, as Isabelle/HOL supports function currying, the inference rule for currying is also provided, i.e.

%Curring
\begin{equation*}
\boxed{
\begin{gathered}
\frac{\sum(F)= \tau_1 \Rightarrow \cdots \Rightarrow \tau_n \Rightarrow \tau_r \quad \Gamma \vdash e_1:\tau_1,\cdots e_i:\tau_i \quad 1\leq i< n}{\Gamma \vdash F \ e_1  \ \cdots  e_i : \tau_{i+1}\Rightarrow\cdots\Rightarrow\tau_n\Rightarrow\tau_r}\quad[\mathbf{Currying}] \\
\end{gathered}
}
\end{equation*}

As we have mentioned above, the type of an expression can be inferred by a top-down approach or a bottom-up approach, and there are situations where more than one type can be derived for one expression due to the using of different inference rules. Solving the inconsistent type information is known formally as type unification. Therefore, a unification mechanism must be introduced to unify different type expressions that assigned to one expression. Briefly, unification is the process of finding substitutions for type variables to make the different type expressions identical. 

In this paper, a substitution is represented by giving the relevant type variable a binding to some type expression.
To obtain all substitutions, the binary comparison abstract-concrete relationship $\succeq$ (and $\succ$) is inductively defined:

\begin{defi}{The abstract-concrete relationship $\succeq$}
\begin{itemize}
  \item $if\ \forall \alpha,\ \beta \in TV,\ then\ \alpha \succeq \beta,\ \beta \succeq \alpha$;
  \item $if\ \forall \alpha \in TV,\ T \in BT,\ then\ \alpha \succ T$;
  \item $if\ \tau \succeq \sigma,\ then\ (\tau) \succeq (\sigma)$;
  \item $if\ \tau_1 \succeq \sigma_1,\ then\ (\tau_1,\ \tau_2) \succeq (\sigma_1,\ \sigma_2)$;
  \item $if\ \tau_1 \succeq \sigma_1,\ \dots\ \tau_n \succeq \sigma_n,\ then\ \tau_1 \ \dots\ \tau_n\ D \succeq \sigma_1\ \dots\ \sigma_n\ D$;
  \item $if\ \tau_1 \succeq \sigma_1,\ \dots\ \tau_n \succeq \sigma_n,\ then\ \tau_1 \Rightarrow \dots \Rightarrow \tau_n \succeq \sigma_1 \Rightarrow \dots \Rightarrow \sigma_n $.
\end{itemize}
\end{defi}

Here, $BT$ is the set of all primitive types (such as $nat, bool, etc.$), and $TV$ is the set of all type variables (such as $\alpha$,$\beta$); $D$ is an n-ary abstract type constructor where $\tau_i$s are its type parameters. 
In above definition, the $\succeq$ can be changed to $\succ$ once one of its condition have one or more $\succ$ holds. It should be noted that, the “abstract-concrete” relationship is different from the concept of subtype, as it is undecidable to say which variable is more “abstract” or “concrete” and both are acceptable for computation. Therefore, “$\alpha \Rightarrow \alpha$” is a subtype of “$\alpha\Rightarrow\beta$”, but in our definition of “abstract-concrete” relationship, both “$\alpha\Rightarrow\alpha\succeq\alpha\Rightarrow\beta$” and “$\alpha\Rightarrow\beta\succeq\alpha\Rightarrow\alpha$” are correct according to this definition.

If a type $\tau$ is more abstract than $\sigma$, i.e $\tau\succeq\sigma$, the relationship can be reduced to relationships with more basic types according to the type structure. Thus, the concept of reduction, denoted by $\Longrightarrow$, is defined:\\

\begin{defi}{Reduction}
\begin{itemize}
  \item $(\tau) \succeq (\sigma) \Longrightarrow \tau \succeq \sigma$;
  \item $(\tau_1,\ \tau_2) \succeq (\sigma_1,\ \sigma_2)\ \Longrightarrow \tau_1 \succeq \sigma_1,\ \tau_2 \succeq \sigma_2$;
  \item $\tau_1 \ \dots\ \tau_n\ D \succeq \sigma_1\ \dots\ \sigma_n\ D \Longrightarrow \tau_1 \succeq
\sigma_1,\ \dots\ \tau_n \succeq \sigma_n$;
  \item $\tau_1 \Rightarrow \dots \Rightarrow \tau_n \succeq \sigma_1 \Rightarrow \dots \Rightarrow \sigma_n \Longrightarrow \tau_1 \succeq \sigma_1,\ \dots\ \tau_n \succeq \sigma_n$.
\end{itemize}
\end{defi}

Reduction is an inference process on the concrete-abstract relationship. For a specific relationship $\tau \succeq \sigma$, we can apply reduction multiple times, making the result irreducible. It can be proved inductively that for an abstract-concrete relationship $\succeq$, the left of the final reduction result must be a type variable, i.e. if $\tau \succeq \sigma$ can be finally reduced to $\alpha \succeq \delta$ where $\alpha$ is a type variable, $\alpha \succeq \delta$ cannot be reduced anymore. Then $\tau \succeq \sigma \overset{*} \Longrightarrow \alpha \succeq \delta$ is used to denote the whole reducing process and $\alpha := \delta$ is defined as its type substitution. Since reduction may produce branches, the final result of reduction can be represented as a set of type substitutions. In this paper, $\mathcal{S}$ is used to denote the type substitution set, i.e. $\mathcal{S}=\{\alpha_1:= \tau_1,\dots,\ \alpha_n:= \tau_n\}$. 

Based on the concepts of abstract-concrete relation and reduction, the rule of type unification is defined as follows:

\begin{equation*}
\boxed{
\begin{gathered}
\frac{\Gamma \vdash e:\tau \quad \Gamma \vdash e:\sigma \quad \tau \succeq \sigma \quad \tau \succeq \sigma \overset{*}\Longrightarrow \mathcal{S}}{\Gamma = \mathcal{S} \Gamma\ } \quad [\mathbf{Uni}]
\end{gathered}
}
\end{equation*}

In this rule, $\mathcal{S}\Gamma$ means making type variable replacement of $\Gamma$ with respect to $\mathcal{S}$.
The scenarios for type unification are different, and the application of type consistency rules is also different. According to the scenario of type unification, the rule $\mathbf{Uni}$ can be specified into the unification rules for application expressions and lambda expressions respectively:

\begin{small}
\begin{equation*}
\boxed{
\begin{gathered}
\frac{\Gamma \vdash F \ e_1 \ \cdots e_n : \tau_r\quad \sum(F)= \tau_1 \Rightarrow \cdots \Rightarrow \tau_n \Rightarrow \tau_r \quad \Gamma \vdash e_i:\sigma_i \quad \Gamma \vdash e_i:\tau_i}
{\sigma_i \succeq \tau_i \overset{*}\Longrightarrow \mathcal{S},\ \Gamma = \mathcal{S} \Gamma\ if\ \sigma_i \succeq \tau_i,\ \tau_i \succeq \sigma_i \overset{*}\Longrightarrow \mathcal{S},\ \Gamma = \mathcal{S} \Gamma\ if\ \tau_i \succeq \sigma_i}
\quad[\mathbf{Uni-App}] \\
\frac{\parbox{12cm}{\hphantom{aaa}$\Gamma,\ x_1:\tau_1, \dots, x_n:\tau_n \vdash F \ e_1 \ \cdots  e_n : \tau_r\quad \sum(F)= \sigma_1 \Rightarrow \cdots \Rightarrow \sigma_n \Rightarrow \sigma_r \\
 \hphantom{aaaaaaaaaaaaaaaaaaaaaaaaa} e_i=x_j\quad 1\leq i,j\leq n$}}
{\parbox{12cm}{$\Gamma \vdash \lambda x_1 \dots x_j \dots x_n.~F \ e_1 \ \cdots  e_n:\tau_1 \Rightarrow \cdots \tau_{j-1} \Rightarrow \sigma_i \Rightarrow \tau_{j+1} \cdots \Rightarrow \tau_n \Rightarrow \tau_r$}}
\quad[\mathbf{Uni-Abs}] \\
\end{gathered}
}
\end{equation*}
\end{small}

\section{Type Inference of Isabelle2Cpp}\label{S:five}
The type inference system of Isabelle2Cpp includes three modules [see Fig.4]: the pattern parameter type extraction module, the bottom-up type inference module and the top-down type completion module.

\begin{figure}[H]
\centerline
  {\includegraphics[scale=0.5]{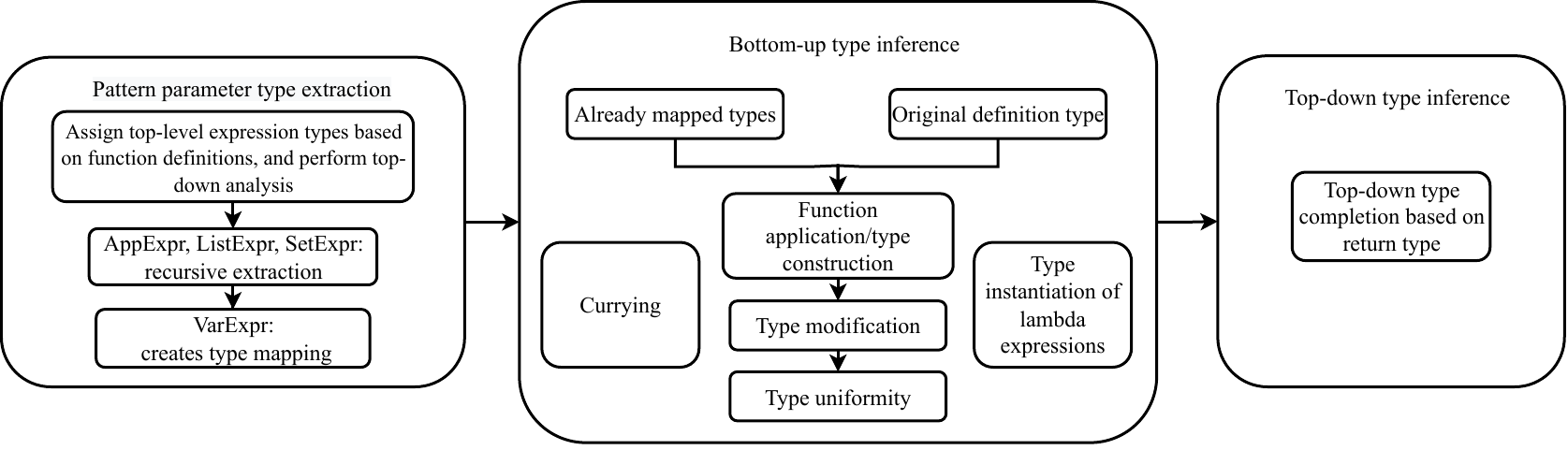}}\vskip3mm

\caption{\label{fig:4} Modules of the type system.}

\end{figure}

\subsection{Pattern Parameter Type Extraction Module}

In a function specification of Isabelle/HOL, an expression on the right side of an equation may be composed of existing functions, datatype constructors, and pattern parameters that appears on the left side of the same equation. As the types of existing functions and datatype constructors can be found in previous defined specifications, which are recorded by the type solver $\Sigma$, to infer the types of the expression and its sub-expressions on the right side, it only needs to extract the type information of all pattern parameters on the left side.

The pattern parameter type extraction module aims to obtain the types of all pattern parameters. In a function specification, a pattern on the left side of an equation is composed of a datatype constructor and the relevant parameters or sub-patterns. As the type of the pattern is determined by the type of corresponding function parameter, the types of its pattern parameters (or sub-patterns) can be inferred by the top-down rules accordingly. The algorithm for pattern parameter type extraction is as follows:

\begin{algorithm}[H]
\renewcommand{\thealgorithm}{1}
\caption{$EX(\Gamma,\ e)$} 
\begin{algorithmic}[1]

\If {$e \in AppExpr$}
    \State $\sum(e.constructor)=\tau_1 \Rightarrow \tau_2 \cdots \Rightarrow \tau_n \Rightarrow \tau_r$
    \For{$i = 1\ to\ e.args\_num$}
        \State $\Gamma = \Gamma \cup \{e.arg_i: \tau_i\}$
        \State $EX(\Gamma,\ e.arg_i)$
    \EndFor

    \ElsIf{$e \in ListExpr$}
        \State $\Gamma \vdash e:\tau\ list$
        \For{$i = 1\ to\ e.exprs\_num$}
            \State $\Gamma = \Gamma \cup \{e.expr_i: \tau\}$
            \State $EX(\Gamma,\ e.expr_i)$
        \EndFor

    \ElsIf{$e \in SetExpr$}
        \State $\Gamma \vdash e:\tau\ set$
        \For{$i = 1\ to\ e.exprs\_num$}
            \State $\Gamma = \Gamma \cup \{e.expr_i: \tau\}$
            \State $EX(\Gamma,\ e.expr_i)$
        \EndFor

    \ElsIf{$e \in VarExpr$}
        \State \Return

    \ElsIf{$e \in ConstExpr$}
        \State \Return

    \Else \State $\Gamma = \Gamma \cup \{e:\bot\}$
    \EndIf
\end{algorithmic}
\end{algorithm}

In the algorithm of pattern parameter type extraction $EX$, the input $\Gamma$ is the environment that includes all type information of the function parameters, and $e$ is a pattern expression to be processed. In function specification, an pattern expression $e$ is either a variable expression ($VarExpr$) or a value construction expression ($ConstExpr$ or $AppExpr$). If the current pattern expression is $VarExpr$ or $ConstExpr$, there is no subexpression and the algorithm terminates type extraction for current branch; if it is $AppExpr$, the algorithm will obtain the type of the constructor $e.constructor$ from $\Sigma$ and assign each pattern parameter $e.arg_i$ of $e$ to the type of corresponding parameter of $e.constructor$. 

As a special case of $AppExpr$, $ListExpr$ or $SetExpr$ may also appear in pattern expressions. For such a case, the overall type ($\tau~list$ or $\tau~set$) of the expression must have been derived and stored in the type environment $\Gamma$. Therefore, the algorithm will first obtain the overall type of the expression from $\Gamma$, such as $\tau~list$ or $\tau~set$, and then use $\mathbf{List-TD}$ rule or $\mathbf{Set-TD}$ rule to obtain the type of its elements. For other kind of expressions, the wrong type $\bot$ will be marked.

The algorithm will process recursively until the types of all pattern expression and variables are inferred.
After the processing, all variables and their corresponding type information will be stored in the $\Gamma$, which is stored in the improved AST.

\subsection{Bottom-Up Type Inference Module}
The bottom-up type inference module is used to infer the types of all expressions on the right side of function equations. It involves three parts: type modification, type unification and the bottom-up type inference. 

\subsubsection{Type modification}
Type modification is to eliminate type variable conflicts between functions and type constructors. 
Type variables are used as type parameters for abstract type construction in specifications. The symbols of type variables in a given specification are replaceable, and they are only used to distinguish different type variables in that specification. If type variables are treated as fixed symbols, type conflicts will occur during type inference. Therefore, this paper introduces type modification to deal with type variable conflicts caused by different functions using the same type variable symbols.

Type modification occurs when an expression is assigned its original defined type. For an expression with the original defined type, the name of the type variable in the expression type will be modified to distinguish the expression type from its original defined type. Specifically, the method of increasing the count number is used to complete the distinction. For example, the original defined type of the map function is $({'d} \Rightarrow {'e})\Rightarrow {'d}\  list \Rightarrow {'e}\ list$, $Nil$, and the original defined type of the $Nil$ constructor is ${'a}\ list$. When assigning types, the original defined type of the $map$ will be modified to $({'d}\#\_ \Rightarrow {'e}\#\_) \Rightarrow {'d}\#\_ \ list \Rightarrow {'e}\#\_ \ list$, the original defined type of $Nil$ will be modified to ${'a}\#\_ \ list$. Among them \_ represents a counter value, which will be incremented after a decoration action, so as to avoid symbol conflicts caused by assignment of functions or constructor types.

Note that the function's type may also come from the left-hand pattern. In this case, we consider that the variable with the function type appearing on the left side has the actual type and does not modify its type.

\subsubsection{Type unification}
Type unification is process of finding the type substitutions that makes the inferred types of one expression equal.
In the process of type inference, $\Gamma$ is the environment which stores the obtained type information for expressions. Due to differences of type information sources and rules for inference, an expression may have different typings in $\Gamma$. In this case, it is necessary to unify the types of the expression.
In this paper, function application expressions and function abstraction expressions, i.e. $AppExpr$ and $LambdaExpr$, are considered for type unification. The unification algorithm is implemented as follows:

\begin{breakablealgorithm}
\renewcommand{\thealgorithm}{2}
\caption{$Unify(\Gamma,\ e)$}
\begin{algorithmic}[1]
    \If{$e \in AppExpr$}
        \State $\sum(e.constructor)=\tau_1 \Rightarrow \cdots \Rightarrow \tau_n \Rightarrow \tau_r$
        \For {$i = 1\ to\ n$}
            \If {$\Gamma \vdash e_{arg_i}:\tau_i\ and\ \Gamma \vdash e_{arg_i}:\sigma_i$}
                \If {$\sigma_i \succeq \tau_i$}
                    \State {$ \sigma_i \succeq \tau_i \overset{*}\Longrightarrow \mathcal{S}_i $}
                    \State {$ \Gamma = \mathcal{S}_i \Gamma $}
                \Else \State $\mathtt{\textbf{continue}}$
                \EndIf
            \EndIf
        \EndFor
    \ElsIf{$e \in LambdaExpr$}
        \If{$e.expr \in AppExpr$}
        \State $\sum(e.expr.contructor)= \sigma_1 \Rightarrow  \cdots \Rightarrow \sigma_n \Rightarrow \sigma_r$
        \State $\Gamma,\ e.parameter_1:\tau_1,\ \dots\ e.parameter_n:\tau_n \vdash e.expr : \tau_r\quad $
            \For{$1\leq i,\ j\leq n$}
                \If {$e_i = x_j$}
                    \State $\tau_j = \sigma_i$
                    \State $\tau_e = \tau_1 \Rightarrow \cdots \Rightarrow \tau_{j-1} \Rightarrow \sigma_i \Rightarrow \tau_{j+1} \cdots \Rightarrow \tau_n \Rightarrow \tau_r$
                \EndIf
            \EndFor
        \Else \State \Return
        \EndIf
    \Else \State \Return
    \EndIf
\end{algorithmic}
\end{breakablealgorithm}

For a function application expression, the types of its parameters can be obtained from two sources: one is derived from the bottom-up type inference for the parameter, and the other is obtained from the function type (or constructor type) according to the position of the parameter in the application expression. This may lead one parameter expression has two different type expressions. Once this happens, the algorithm will perform one or more reductions based on the abstract-concrete relationship $\succeq$ of the two types, and use the obtained substitution set $\mathcal{S}_i$ for the replacement of corresponding type variables in $\Gamma$. After the replacement, the types of the function application expression and its parameters will be unified.

For a function abstraction expression, i.e. a lammbda-expression, the symbol $\lambda$ introduces a set of $\lambda$-parameters binding for the $\lambda$-body. When the $\lambda$-body is a function application (or a value construction), a $\lambda$-parameter directly appeared in the function application (or the value construction) also has two sources of type information, i.e. the type introduced by $\lambda$-parameter and the type inferred from the function application. Thus, the types of parameters and the type of the whole lambda-expression all need to be unified.

\subsubsection{The bottom-up type inference}
The type system of Isabelle2Cpp uses a bottom-up approach to infer the types of expressions on the right side of function equations. In this process, the bottom expressions of type deduction is either a constant expression or a variable expression, i.e. $ConstExpr$ or $VarExpr$, which can be obtained directly or through pattern parameter type extraction. For a complex expression, once all the types of its sub-expression are available, the bottom-up inference rules can be used to obtain its type. The bottom-up type inference algorithm is provided as follows:

\begin{breakablealgorithm}
\renewcommand{\thealgorithm}{3}
\caption{$BU(\Gamma,\ e)$}
\begin{algorithmic}[1]
    \If{$e \in AppExpr$}
        \For{$i = 1\ to\ e.args\_num$}
            \State $BU(\Gamma,\ e.arg_i)$
        \EndFor
        \State $\sum(e.constructor) = \tau_1 \Rightarrow \cdots \Rightarrow \tau_n \Rightarrow \tau_r$
        \State $\Gamma = \Gamma \cup \{e : F\ \tau_{e.arg_1}\ \tau_{e.arg_2}\ \cdots\ \tau_{e.arg_{args\_num}}\}$
        \State $Unify(\Gamma,\ e)$

    \ElsIf{$e \in ListExpr$}
        \For{$i = 1\ to\ e.exprs\_num$}
            \State $BU(\Gamma,\ e.expr_i)$
        \EndFor
        \State $\Gamma = \Gamma \cup \{e : \tau_{e.expr_1}\ list\}$

    \ElsIf{$e \in SetExpr$}
        \For{$i = 1\ to\ e.exprs\_num$}
            \State $BU(\Gamma,\ e.expr_i)$
        \EndFor
        \State $\Gamma = \Gamma \cup \{e : \tau_{e.expr_1}\ set\}$

    \ElsIf{$e \in LetInExpr$}
        \State $BU(\Gamma,\ e.equation.expr)$
        \State $\Gamma = \Gamma \cup \{e.equation.pattern : \tau_{e.equation.expr}\}$
        \State $EX(\Gamma,\ e.equation.pattern)$
        \State $BU(\Gamma,\ e.expr)$
        \State $\Gamma = \Gamma \cup \{e : \tau_{e.expr}\}$

    \ElsIf{$e \in CaseExpr$}
        \State $BU(\Gamma,\ e.expr)$
        \For{$i = 1\ to\ e.equations\_num$}
            \State $\Gamma = \Gamma \cup \{e.equation_i.pattern : \tau_{e.expr}\}$
            \State $EX(\Gamma,\ e.equation_i.pattern)$
            \State $BU(\Gamma,\ e.equation_i.expr)$
        \EndFor
        \State $\Gamma = \Gamma \cup \{e : \tau_{e.equation_1.expr}\}$

    \ElsIf{$e \in LambdaExpr$}
        \For{$i = 1\ to\ e.parameters\_num$}
            \State $\Gamma = \Gamma \cup \{e.patameter_i:\tau_{lambda@lambda\_counter}\}$
        \EndFor
        \State $BU(\Gamma,\ e.expr)$
        \State $\Gamma = \Gamma \cup \{e : \tau_{e.parameter_1} \Rightarrow \cdots \Rightarrow \tau_{e.parameter_{e.parameters\_num}} \Rightarrow \tau_{e.expr}\}$
        \State $Unify(\Gamma,\ e)$

    \ElsIf{$e \in VarExpr$}
        \State \Return

    \ElsIf{$e \in ConstExpr$}
        \If{$e \in IntegralExpr$}
            \State $\Gamma = \Gamma \cup \{e:nat\}$
        \Else
            \State $\Gamma = \Gamma \cup \{e:bool\}$
        \EndIf

    \Else \State $\Gamma = \Gamma \cup \{e:\bot\}$
    \EndIf
\end{algorithmic}
\end{breakablealgorithm}

In the algorithm $\mathbf{BU}$, type inference is performed recursively: if an expression is composed by several subexpressions, $\mathbf{BU}$ is used to infer the types of all its subexpressions; the type information of subexpressions is stored in $\Gamma$, and the relevant bottom-up rules are used to complete the type inference of the whole expression. In addition, as type inconsistencies may occur in the derivation of $AppExpr$ and $LambdaExpr$, the unification algorithm $Unify$ is applied to correct the types of the expressions and relevant parameters. If an expression is $Const$ or $VarExpr$, its type has already been obtained by the pattern parameter type extraction module and stored in $\Gamma$, so no operation is performed. If an expression does not belong to one of these types, it is marked as the wrong type.
Once the types of all expressions on the right side of the function equations have been inferred, all type information are stored in the improved AST.

\subsection{Top-Down Type Completion Module}
After the bottom-up type inference module, most of the type missing problems can be solved, but there are still problems with the type confirmation for some expressions. Assuming that the type of the overall expression on the right side of a function equation is $\tau_X$, the type may be inconsistent with the type $\tau_N$ of the return value provided in the function specification. This is because some of the terms in the expression have their originally defined types and the process of expression construction can not provide further type information for their instantiation. This will result in the possibility that the type of the final expression is not fully instantiated. 
Take the following specification $product\_lists$ as an example:

\begin{Verbatim}[frame=single, fontsize=\small]
primrec product_lists :: "'a list list => 'a list list" where
  "product_lists [] = [[]]" |
  "product_lists (xs # xss) = concat (map (\<lambda>x.map (Cons x)
    (product_lists xss)) xs)"
\end{Verbatim}

In Isabelle/HOL, [] (a symbolic representation of $Nil$) has a original defined type of ${'a}\ list$. Assuming that the type of [] on the right side of the first equation has been modified to be ${'a}\#1\ list$, then the type of the right [[]] will be ${'a}\#1\ list\ list$ according to the bottom-up type inference; However, according to $product\_lists$ The return value type defined by the list function, [[]], should be of type ${'a}\ list\ list$, causing a conflict.

For this case, type unification cannot solve this problem. This is because type unification is processed from subexpressions or parameters to the whole expression. But in above problem, the inconsistent comes from the types of the whole expression but the types of all subexpressions or parameters are currently acceptable. Therefore, after completing the bottom-up type inference, it is necessary to use the type information provided by the function specification to complete the top-down type completion of the right expression again.
The top-down type completion algorithm is provided as follows:

\begin{breakablealgorithm}
\renewcommand{\thealgorithm}{4}
\caption{$TD(\Gamma,\ e)$}
\begin{algorithmic}[1]
\If{$e \in AppExpr$}
    \State $\sum(e.constructor)=\tau_1 \Rightarrow\cdots \Rightarrow \tau_n \Rightarrow \tau_r$
    \State {$\Gamma \vdash e:\tau_e$}
        \If {$\tau_r \succeq \tau_e\ and\ $}
            \State {$ \tau_r \succeq \tau_e \overset{*}\Longrightarrow \mathcal{S} $}
        \Else \State {$ \mathcal{S} = [\ ]$}
        \EndIf

    \For{$i = 1\ to\ e.args\_num$}
        \If {$\tau_{e.arg_i} \succeq \mathcal{S}\tau_i$}
            \State {$\tau_{e.arg_i} = \mathcal{S}\tau_i$}
        \EndIf
        \State $TD(\Gamma,\ e.arg_i)$
    \EndFor

    \ElsIf{$e \in ListExpr$}
        \State $\Gamma \vdash e:\tau\ list$
        \For{$i = 1\ to\ e.exprs\_num$}
            \State $\Gamma = \Gamma \cup \{e.expr_i: \tau\}$
            \State $TD(\Gamma,\ e.expr_i)$
        \EndFor

    \ElsIf{$e \in SetExpr$}
        \State $\Gamma \vdash e:\tau\ set$
        \For{$i = 1\ to\ e.exprs\_num$}
            \State $\Gamma = \Gamma \cup \{e.expr_i: \tau\}$
            \State $TD(\Gamma,\ e.expr_i)$
        \EndFor

    \ElsIf{$e \in LetInExpr$}
        \State $\Gamma \vdash e:\tau$
        \State $\Gamma = \Gamma \cup \{e.expr: \tau\}$
        \State $TD(\Gamma,\ e.expr)$

    \ElsIf{$e \in CaseExpr$}
        \State $\Gamma \vdash e:\tau$
        \For{$i = 1\ to\ e.equations\_num$}
            \State $\Gamma = \Gamma \cup \{e.equation_i.expr: \tau\}$
            \State $TD(\Gamma,\ e.equation_i.expr)$
       \EndFor

    \ElsIf{$e \in LambdaExpr$}
        \State \Return

    \ElsIf{$e \in VarExpr$}
        \State \Return

    \ElsIf{$e \in ConstExpr$}
        \State \Return

    \Else \State $\Gamma = \Gamma \cup \{e:\bot\}$
    \EndIf
\end{algorithmic}
\end{breakablealgorithm}

In the algorithm $TD$, the type of the whole expression on the right side of each equation will be assigned to the type of the return value according to the specification, and type inference of its sub-expressions is completed in sequence. For each expression, the algorithm needs to determine the abstract-concrete relationship between the type obtained by the top-down type completion and the type obtained by the bottom-up inference: if the former is more abstract than the latter, type replacement needs to be completed, and then its subexpressions need to be processed accordingly. After completing the top-down type completion, the algorithm will update the type information in the improved AST.

It should be noted that top-down type completion should to be processed after the bottom-up type inference. This is because the types of some expressions cannot be inferred from the types of their sup-expressions. For example, the expression $expr_1=expr_2$ has an type of $bool$, but we cannot obtain the types of its left and right subexpressions from the type bool. There is no constraint relationship between its sub-expression types and the expression type, and it is mostly a polymorphic operator defined by class (such as$=, >$, etc.). Therefore, if $AppExpr$ is a polymorphic operator such as $'='$, it cannot complete top-down type completion, and the current result needs to be returned directly.

\section{Case Study}\label{S:six}
The type system proposed in this paper solves the problems of type unification and type inference of lambda expressions in the previous Isabelle2Cpp framework. With this type system, the Isabelle2Cpp generation framework can automatically obtain more complete type information, and improve the process of C++ code generation. This chapter will illustrate the effect of Isabelle2Cpp type inference by comparing the generated C++ codes from the specifications of binary search $bs$ and $product\_lists$ by using the type system.

In Isabelle/HOL, the specification $bs$ for binary search is defined as follows:

\begin{Verbatim}[frame=single, fontsize=\small]
fun bs :: "nat => nat list => nat option" where
  "bs x [] = None" |
  "bs x [y] = If (x = y) (Some 0) None" |
  "bs x ys = (let m = (length ys) div 2 in
      let y = ys ! m in
        If (y = x)
          (Some m)
          (If (y < x)
            (case bs x (drop (m + 1) ys) of Some n => Some
            (m + n + 1) |
                None => None)
            (bs x (take m ys)
          )
      )
  )"
\end{Verbatim}

After using the type inference algorithms proposed in this paper, the C++ codes generated by Isabelle2Cpp from $bs$ is as follows. Statements generated before using the type inference system proposed in this article are marked with '-' at the beginning of the sentence, and statements generated after using the type system proposed in this article are marked with '+' at the beginning of the sentence:

\begin{Verbatim}[frame=single, fontsize=\small]
#include "binary_search.hpp"

std::optional<std::uint64_t> bs(const std::uint64_t &arg1,
                                std::deque<std::uint64_t> arg2) {
    // bs x [] = None
    if (arg2.empty()) {
        return std::optional<std::uint64_t>();
    }

    // bs x [y] = If (x = y) (Some 0) None
    if (arg2.size() == 1) {
        auto y = arg2[0];
        std::optional<std::uint64_t> temp0;
        if (arg1 == y) {
            temp0 = std::make_optional<std::uint64_t>(0);
        } else {
            temp0 = std::optional<std::uint64_t>();
        }
        return temp0;
    }

    // bs x ys = (let m = (length ys) div 2 in ...
    auto temp0 = arg2.size() / 2;
    auto m = temp0;
    auto temp2 = arg2;
    auto temp1 = temp2[m];
    auto y = temp1;
    std::optional<std::uint64_t> temp3;
    if (y == arg1) {
        temp3 = std::make_optional<std::uint64_t>(m);
    } else {
        std::optional<std::uint64_t> temp4;
        if (y < arg1) {
            auto temp5 = ([&] {
-               auto temp6 = bs(arg1, decltype(arg2)(arg2.begin() +
                            m + 1, arg2.end()));
+               auto temp6 = bs(arg1, std::deque<std::uint64_t>(arg2.begin() +
                            m + 1, arg2.end()));
                    ...
            })();
            temp4 = temp5;
        } else {
-           temp4 = bs(arg1, decltype(arg2)(arg2.begin(),
                    arg2.begin() + m));
+           temp4 = bs(arg1, std::deque<std::uint64_t>(arg2.begin(),
                    arg2.begin() + m));
        }
        temp3 = temp4;
    }
    return temp3;
}
\end{Verbatim}

For the generated codes without the type system, the type of the second parameter of $bs$ is specified according to the type of $arg2$ by $\mathbf{decltype}$ in C++; but with type system, Isabelle2Cpp can directly obtain std::deque$<$std::uint64\_t$>$ as the type of the parameter. It can be seen that by using the type system proposed in this article, the dependence of the Isabelle2Cpp code generation framework on  $\mathbf{auto}$ and  $\mathbf{decltype}$ in C++ can be eliminated, thereby generating code with richer type information.

In the above example of $product\_lists$, the type of [[]] and the type of $\lambda~x.~map~(Cons~x)$\ $(product\_lists~ xss)$ cannot be inferred through the type inference of C++; besides, the code generation of this function also requires to obtain the type of the curried expression $Cons~x$. With the help of the type system proposed in this paper, it can be automatically inferred that the type of [[]] is $'a\ list\ list$, which is consistent with the type of the return value of  $product\_lists$, and the type of $(Cons\ x)$ is inferred as the function type $'a\ list \Rightarrow 'a\ list$, which makes the type of the lambda-expression finally marked as $'a \Rightarrow {'a}\ list\ list$. The specific type inference results are shown below. In the type inference results shown below, we have typed the expressions at each granularity in the $product\_lists$ function definition. We use parentheses to combine an expression and its type, and use :: to split the expression and its corresponding type.

\begin{Verbatim}[frame=single, fontsize=\small]
product_lists
([(Nil :: ('a )list)] :: (('a )list )list)
((concat ((map (\<lambda>x.((map ((Cons (x :: 'a)) :: (('a )list => ('a )
list))((product_lists (xss :: (('a )list )list)) :: (('a )list)list))
 :: (('a )list )list):: ('a => (('a )list )list))(xs :: ('a)list)) ::
  ((('a )list )list )list)):: (('a )list )list)
\end{Verbatim}

Through the above cases, it can be seen that the type system proposed in this paper eliminates the dependence of Isabelle2Cpp on the type inference in C++, it can provide type inference for pattern parameters and all kinds of expressions, which provides effective support for the code generation of Isabelle2Cpp.
\section{Conclusion}\label{S:seven}

This paper proposes a type system for the code generation framework Isabelle2Cpp, which is used to do type inference and type unification for expressions of the intermediate representation. The system includes three modules that cooperate with each other: the pattern parameter type extraction module, the bottom-up type inference module, and the top-down type completion module. The pattern parameter type extraction module is used to obtain the type information of variables other than functions and data type constructors, providing a reliable source of type information. The bottom-up type inference module obtains the detailed type information of all expressions and complete the type unification. The top-down type completion module integrates the type information obtained by the bottom-up type inference with the type information provided by the function specification, and completes the type consistency checking of the whole expression. By incorporating the type system, Isabelle2Cpp provides more comprehensive type information for expression generation, and can automatically generate the corresponding C++ codes even if some type information is missing in Isabelle/HOL specification.

\section*{Acknowledgement}
{\rm This work was supported by Beijing Excellent Talents Subsidized Project.}

\bibliographystyle{alphaurl}
\bibliography{lmcs-bibliography}

\end{document}